\documentclass[a4paper,11pt]{article}

\pdfoutput=1
\usepackage{jheppub}

\usepackage[T1]{fontenc}

\usepackage{bm}
\usepackage{array}
\usepackage{multirow}
\usepackage{booktabs}
\usepackage[normalem]{ulem}
\usepackage{makecell}

\usepackage{multicol}
\usepackage{mleftright}
\usepackage[utf8]{inputenc}
\usepackage{slashed}
\usepackage{abraces}
\usepackage{mathtools}
\usepackage{amssymb}
\usepackage{amsmath}
\usepackage{bbold}
\usepackage{slashed}
\usepackage{cancel}
\usepackage{color}
\usepackage[dvipsnames]{xcolor}
\usepackage{epsfig}

\usepackage{hhline}
\newcommand\Tstrut{\rule{0pt}{3.2ex}}
\newcommand\Bstrut{\rule[-1.8ex]{0pt}{0pt}}

\usepackage{tikz}
\usepackage{pgfplots}
\pgfplotsset{compat=1.15}
\usepgfplotslibrary{polar}
\usepgflibrary{shapes.geometric}
\usetikzlibrary{calc}
\graphicspath{{}}
\usepackage{float} %For Tables
\usepackage{feynmp-auto} %Feynman Diagrams
\usepackage{tensor} %Indices placement for tensors
\usepackage{physics} %Bunch of commands for equations
\usepackage[caption=false]{subfig}
\usepackage{listings}
\usepackage{url}

\renewcommand{\i}{\mathrm i}

\renewcommand{\[}{\left[}

\definecolor{bostonuniversityred}{rgb}{0.8, 0.0, 0.0}

\newcommand{\redBU}[0]{\color{bostonuniversityred}}

\usepackage{pifont}% http://ctan.org/pkg/pifont
\title{Phenomenology at the Large Hadron Collider with Deep Learning: the case of vector-like quarks decaying to light jets}

\author[a]{Felipe F.~Freitas,} 
\emailAdd{felipefreitas@ua.pt}

\author[a]{João Gon\c calves,} 
\emailAdd{jpedropino@ua.pt}

\author[a]{Ant\'onio P. Morais,} 
\emailAdd{aapmorais@ua.pt}

\author[b]{Roman Pasechnik} 
\emailAdd{Roman.Pasechnik@thep.lu.se}

\affiliation[a]{Departamento de F\'\i sica da Universidade 
de Aveiro and Centre  for  Research  and  Development \\  in  Mathematics  and  Applications (CIDMA), Campus de Santiago, 3810-183 Aveiro, Portugal}
	
\affiliation[b]{Department of Astronomy and Theoretical Physics, Lund University,\\ S\"{o}lvegatan 14A, SE 223-62 Lund, Sweden}

\keywords{Beyond the Standard Model, Vector-Like Quarks, Large Hadron Collider, Deep Learning, Machine Learning algorithms}

\abstract{In this work, we continue our exploration of TeV-scale vector-like fermion signatures inspired by a Grand Unification scenario based on the trinification gauge group. A particular focus is given to pair-production topologies of vector-like quarks (VLQs) at the LHC, in a multi-jet plus a charged lepton and a missing energy signature. We employ Deep Learning methods and techniques based in evolutive algorithms that optimize hyper-parameters in the neural network construction, whose objective is to maximise the Asimov estimate for distinct VLQ masses. In this article, we consider the implications of an innovative approach by simultaneously combining detector images (also known as jet images) and tabular data containing kinematic information from the final states. With this technique we are able to exclude VLQs, that are specific for the considered model, up to a mass of  800 GeV in both the high-luminosity the Run-III phases of the LHC programme.}

\begin{document}

\maketitle

%%%%%%%%%%%%%%%%%%%%%%%%%%%%%%%%%%%%%%%%%%%%%%%%%%%%%
\section{Introduction}\label{sec:Introduction}
%%%%%%%%%%%%%%%%%%%%%%%%%%%%%%%%%%%%%%%%%%%%%%%%%%%%%

The Standard Model (SM) of particle physics has, so far, been recognized as the most successful model in theoretical physics, proving its validity over a wide range of subatomic phenomena and energy scales \cite{ATLAS:2012yve,D0:1995jca,CDF:1995wbb,GargamelleNeutrino:1973jyy,UA1:1983crd,UA1:1983mne,UA2:1983mlz}. However, despite its immense success, there is an experimental evidence that indicates certain missing features for which the SM can not provide a satisfactory explanation. Among these are the lack of a dynamical mechanism behind neutrino mass generation and the absence of a plausible dark matter (DM) candidate, as well as the emergence of slowly but steadily growing anomalies such as those of the magnetic moment of the muon \cite{Bennett_2006,PhysRevLett.126.141801}, the $\mathrm{R_{D,D^*}}$ \cite{LHCb:2017smo,PhysRevLett.118.211801} and the $\mathrm{R_{K,K^*}}$ flavour observables \cite{BELLE:2019xld,LHCb:2021trn}. Regarding the muon $g-2$ and B-physics anomalies, one should stress that the corresponding deviations for the SM predictions do not pass the $5\sigma$ threshold for discovery, and thus are only considered as an indication for possible New Physics (NP) effects.

To address these shortcomings one typically extends the SM framework by introducing new beyond-the-SM (BSM) fields and new interactions. There are three potential avenues for one to follow in this case. One possibility relies in a phenomenology driven approach, and either simply add new BSM fields that address a given problem \cite{PhysRevLett.127.061802,Dorsner:2019itg}, or work under the effective field theory (EFT) formalism attempting to establish a bound on the NP energy scale by confronting the model predictions with the data \cite{Capdevila:2017bsm,Buttazzo:2017ixm}. One can also adopt a more fundamental avenue and build a high-energy scale framework from which the SM gauge group, its Yukawa sector and particle content are emergent at low energy-scale. In the latter approach, one typically resorts to the so-called Grand Unified Theories (GUTs), potentially providing a first-principles understanding of the origins of the specific SM structure. For such a class of models, one attempts to encapsulate the electroweak (EW) symmetry group, $\mathrm{SU(3)_C\times SU(2)_L \times U(1)_Y}$ into a larger gauge group, with the SM and NP fields being embedded into its representations. This strategy may result in various types of unification in the sectors of matter fields and forces such as gauge and/or Yukawa couplings unification, Higgs-lepton unification, quark-lepton unification etc, in some cases even yielding relatively light NP fields in the low-energy SM-like EFT spectrum. The latter may be observable at current or near-future collider experiments, offering a potential of probing the GUT-scale physics through either direct observation of those new states or studies of their impact on precision SM observables, such as the muon $g-2$ and flavour changing neutral current (FCNC) processes. In this context, some of the authors have previously constructed a novel class of GUT models based on the trinification gauge group supplemented with a horizontal family symmetry in both supersymmetric (SUSY) and non-SUSY versions \cite{Camargo-Molina:2016bwm,Camargo-Molina:2016yqm,Camargo-Molina:2017kxd,Morais:2020odg,Morais:2020ypd,CarcamoHernandez:2020owa}. In particular, it has been demonstrated that successful unification of the gauge couplings can be achieved in a SUSY formulation in consistency with proton decay constraints \cite{Morais:2020odg,Morais:2020ypd}. For this work, we shall focus on the latter scenario, which shares many common features with other GUT implementations and thus represents a good playground for collider searches.

The SHUT (\textbf{S}upersymmetric \textbf{H}iggs-\textbf{U}nified \textbf{T}rinification) model \cite{Morais:2020odg,Morais:2020ypd} predicts various types of NP at low energies. The extended fermion sectors include heavy left-handed and sterile right-handed neutrinos as well as vector-like fermions (VLFs), such as EW-singlet vector-like quarks (VLQs) and EW-doublet vector-like leptons (VLLs). The latter can potentially be at the reach of Run-III and the high-luminosity (HL) phases of the Large Hadron Collider (LHC) \cite{Freitas:2020ttd}. An extended scalar sector features a minimum of three Higgs doublets at low scales, required for consistency with the measured Cabibbo–Kobayashi–Mas\-kawa (CKM) quark mixing matrix. 

In this article we continue the phenomenological exploration of SHUT inspired models, focusing on the collider signatures of a down-type VLQs, expected to emerge at a TeV energy scale. Three generations of VLQs are predicted within such a GUT framework with the first two generations being light enough to be potentially accessible at the LHC, with masses of $\mathcal{O}(1)~\mathrm{TeV}$, whereas the 3rd generation VLQ is expected to hold a mass of $\mathcal{O}(100)~\mathrm{TeV}$ i.e. it is too heavy to be probed in collider measurements in the near future. As a particular promising discovery channel, we consider multi-jet signatures with one charged lepton and missing energy, probing the couplings of the VLQ with light quarks. For this, we apply Deep Learning (DL) techniques to help separating signal from background classes. To this end, we rely on two types of data in the training phase, namely, we use the abstract jet images \cite{Alves:2019ppy} and the numerical data containing kinematic information from the final states. To train the Neural Network (NN) on the data, we construct a specific Hybrid Net, which is designed to make use of both data formats. Optimization of the NN hyper-parameters is performed by using an evolution algorithm, which optimises the Asimov metric. For a thorough description of our methods and techniques we refer the reader to our previous work \cite{Freitas:2020ttd}.

The paper is organized as follows. In Sec.~\ref{sec:SHUT} we provide a brief introduction into the SHUT model, providing particular details relevant for our phenomenological analysis. In Sec.~\ref{sec:Search} we present the methodology implemented in the search for VLQs at the LHC, including a full description of the chosen topology as well as a detailed description of the DL algorithms and datasets used for solving the considered classification problem. This section is finalised with the presentation and subsequent discussion of the numerical results. Finally, concluding remarks are given in Sec.~\ref{sec:conclusions}. Additional information is presented in appendix, including the VLQ Feynman rules \ref{app:Feynman-Rules} as well as kinematic and angular distributions \ref{app:Kinematics}.

%%%%%%%%%%%%%%%%%%%%%%%%%%%%%%%%%%%%%%%%%%%%%%%%%%%%%%%%%%%%%%%%%%%%%%%%%%%
\section{VLQ sector of the SHUT model}
\label{sec:SHUT}
%%%%%%%%%%%%%%%%%%%%%%%%%%%%%%%%%%%%%%%%%%%%%%%%%%%%%%%%%%%%%%%%%%%%%%%%%%%

In this section, we begin by briefly introducing the SHUT model, with the focus on the main properties that arise from the GUT description, without delving deep into the mathematical inner workings of this framework. A more detailed description of the considered GUT framework can be found in earlier works e.g.~in Refs.~\cite{Camargo-Molina:2016yqm,Camargo-Molina:2017kxd,Camargo-Molina:2016bwm,Morais:2020odg,Morais:2020ypd}, while a comprehensive discussion of the low-energy limit, including the physical Lagrangian that is used in this work, can be found in Ref.~\cite{Freitas:2020ttd}. In what follows we provide the theoretical background that was taken as a physics case to our study. For completeness of information a more extensive discussion of the model can be found in Appendix~\ref{app:Model}.

The SHUT model is a particularly promising flavoured GUT (F-GUT) scenario whose group-theoretical structure is inspired by the $\mathrm{E_6}$ exceptional gauge group extended with a local $\mathrm{SU(2)_F\times U(1)_F}$ family symmetry, where all matter content, including the Higgs sector, is unified in fundamental (\textbf{27,2}) and (\textbf{27,1}) representations. It then proposes a first-principles' explanation for some of the well-known problems within the SM, such as a common origin of all fundamental gauge forces (with exception of gravity) as well as the observed flavour hierarchies in the SM fermionic sectors. In practical terms, the SHUT model starts off with the trinification gauge symmetry, supplemented with the $\mathrm{SU(2)_F\times U(1)_F}$ family symmetry, whose representations have a common origin from the referred (\textbf{27,2}) and (\textbf{27,1}) multiplets of the $\mathrm{E_6} \times \mathrm{SU(2)_F\times U(1)_F}$ gauge group.

Due to SUSY, the Higgs sector is unified with the charged lepton and neutrinos sectors, thus, sharing common flavor properties with fermions at high energies. The superpotential of the SHUT model reads \cite{Morais:2020ypd}
\begin{equation}
\label{eq:Superpotential}
	\begin{aligned}
		W_3 = \varepsilon_{IJ}\qty(\mathcal{Y}_1\bm{L}^I\bm{Q}^3_\mathrm{L}\bm{Q}_R^J - \mathcal{Y}_2\bm{L}^I\bm{Q}^J_\mathrm{L}\bm{Q}_R^3 + \mathcal{Y}_2\bm{L}^3\bm{Q}^I_\mathrm{L}\bm{Q}_R^J) \,,	\end{aligned}
\end{equation}
where $\varepsilon_{IJ}$ is the two-dimensional Levi-Civita symbol, $\mathcal{Y}_1$ and $\mathcal{Y}_2$ are the two unified Yukawa couplings, while subscripts $\mathrm{L}/\mathrm{R}$ denote $\mathrm{SU(3)_{L/R}}$ superfields. The indices $I,J=1,2$ represent $\mathrm{SU(2)_F}$ doublets, while flavour singlets are denoted with the ``3'' in superscript. These massless superfields form bi-triplets of the trinification symmetry ($\mathrm{SU(3)_R \times SU(3)_L \times SU(3)_C}$) and transform in accordance with the representations shown in Tab.~\ref{tab:charges}. Expanding these fields in terms of $\mathrm{SU(3)}$ indices, we write the superfields in component form as \cite{Morais:2020ypd}
\begin{equation}\label{eq:super_trini}
\begin{aligned}
&\qty(\bm{L}^{i,3})\indices{^l_r} = \begin{pmatrix}
\bm{\mathcal{N}}_\mathrm{R} & \bm{\mathcal{E}}_\mathrm{L} & \bm{e}_\mathrm{L} \\
\bm{\mathcal{E}}_\mathrm{R} & \bm{\mathcal{N}}_\mathrm{L} & \bm{\nu}_\mathrm{L} \\
\bm{e}_\mathrm{R} & \bm{\nu}_\mathrm{R} & \bm{\phi}
\end{pmatrix}^{i,3}, \\
&\qty(\bm{Q}_\mathrm{L}^{i,3})\indices{^x_l} = \begin{pmatrix}
\bm{u}_\mathrm{L}^x & \bm{d}_\mathrm{L}^x & \bm{D}_\mathrm{L}^x
\end{pmatrix}, \\
&\qty(\bm{Q}_\mathrm{R}^{i,3})\indices{^r_x} = \begin{pmatrix}
\bm{u}_{\mathrm{R}_x} & \bm{d}_{\mathrm{R}_x} & \bm{D}_{\mathrm{R}_x}
\end{pmatrix}^{\mathrm{T}\,\,\,\, i,3},
\end{aligned}
\end{equation}
where $l$, $r$ and $x$ are $\mathrm{SU(3)_L}$, $\mathrm{SU(3)_R}$ and $\mathrm{SU(3)_C}$ indices, respectively. In here, we note that the left-handed and right-handed components of the SM-like leptons are embedded in the $\bm{L}$ superfield ($\bm{e}_\mathrm{R}$ and $\bm{e}_\mathrm{L}$). Right-handed neutrinos also reside in this representation (which are embedded inside $\bm{\mathcal{N}}_\mathrm{R}$, $\bm{\nu}_\mathrm{R}$ and $\bm{\phi}$), which helps address the neutrino mass problem via a radiative seesaw mechanism \cite{Camargo-Molina:2017kxd,Morais:2020ypd}. The SM quarks are then found in the $\bm{d}_\mathrm{L,R}/\bm{u}_\mathrm{L,R}$ components of $\bm{Q}_{\text{L,R}}$, whereas the new exotic down-type VLQs reside in the $\bm{D}_\mathrm{L,R}$ components.

\begin{table}[htb!]
\centering
\begin{tabular}{llllll}
\hline
                        & $\mathrm{SU(3)_L}$ & $\mathrm{SU(3)_R}$ & $\mathrm{SU(3)_C}$ & $\mathrm{SU(2)_F}$ & $\mathrm{U(1)_F}$ \\ \hline
$\bm{L}^i$              & $\bm{3}$           & $\bm{\bar{3}}$     & $\bm{1}$           & $\bm{2}$           & $1$               \\
$\bm{L}^3$              & $\bm{3}$           & $\bm{\bar{3}}$     & $\bm{1}$           & $\bm{1}$           & $-2$              \\
$\bm{Q}_{\mathrm{L}}^i$ & $\bm{\bar{3}}$     & $\bm{1}$           & $\bm{3}$           & $\bm{2}$           & $1$               \\
$\bm{Q}_{\mathrm{L}}^3$ & $\bm{\bar{3}}$     & $\bm{1}$           & $\bm{3}$           & $\bm{1}$           & $-2$              \\
$\bm{Q}_{\mathrm{R}}^i$ & $\bm{1}$           & $\bm{3}$           & $\bm{\bar{3}}$     & $\bm{2}$           & $1$               \\
$\bm{Q}_{\mathrm{R}}^3$ & $\bm{1}$           & $\bm{3}$           & $\bm{\bar{3}}$     & $\bm{1}$           & $-2$             
\end{tabular}
\caption{Trinification fields representations under $\mathrm{[SU(3)]^3 \times SU(2)_F \times U(1)_F}$ for each of the fundamental chiral superfields shown in Eq.~\eqref{eq:Superpotential}.}
\label{tab:charges}
\end{table}

At low energies, the SM gauge group is fully realized, with additional exotic physics, including an extended scalar sector and vector-like fermions. Full unification of the field content results in a reduced freedom in the Yukawa sector, with the presence of only two $\mathcal{Y}_1$ and $\mathcal{Y}_2$ couplings at high scales, which provide the leading contributions to the generation of quark masses at tree level and the necessary means to radiatively induce strong hierarchies as observed in Nature. Indeed, as soon as the scalar components of $\bm{L}$ develop vacuum expectation values (VEVs) that break the SHUT symmetry down to that of the SM, the down-type VLQs, second- and third-generation SM-like quarks gain tree-level masses, while the charged leptons and first-generation quarks develop their masses at quantum level via loop corrections. This fact offers the necessary ingredients to potentially explain the observed splitting in the fermion spectra implying the relative lightness of the charged leptons, neutrinos and the quarks of first generation.

In Ref.~\cite{Morais:2020ypd}, various benchmark scenarios assuming three light Higgs doublets yielding a consistent CKM mixing matrix, were explored in detail. The further presence of new VLQs results in an extended $3\times 6$ quark mixing matrix, defined as
\begin{equation}\label{eq:VLQs_CKM}
V_\text{CKM} = U_L^u \dotproduct P \dotproduct (U_L^d)^\dagger = \begin{pmatrix}
V_{\mathrm{CKM}}^{\mathrm{SM}} & V_{\mathrm{CKM}}^{\mathrm{VLQs}}
\end{pmatrix} \,,
\end{equation}
where $P$ is a projection operator, defined as $P = \begin{pmatrix}\mathbb{1}_{3\times 3} \hspace{0.5em} | \hspace{0.5em} 0_{3\times 3}\end{pmatrix}$. Here, a $3\times 3$ matrix $V_{\mathrm{CKM}}^{\mathrm{SM}}$ represents the CKM-like quark mixing of the SM while a $3\times 3$ matrix $V_{\mathrm{CKM}}^{\mathrm{VLQs}}$ contains a mixing between the down-type VLQs with up-type chiral quarks of the SM. Two of them are relatively light, with a certain hierarchy between them, and are expected to emerge at a TeV energy scale. A third heavy VLQ appears beyond the reach of current or near-future colliders. 

In the present work we explore the LHC discovery potential of the lightest VLQ denoted as $D$, ignoring its heavier counterparts for simplicity. Their mass reads as
\begin{equation}
 m_\text{D}\sim \omega\mathcal{Y}_2 \qquad m_\text{S}\sim p\mathcal{Y}_2 \qquad m_\text{B}\sim p\mathcal{Y}_1\,,
\label{eq:mVLQ}
\end{equation}
with $0.01 \sim \mathcal{Y}_2 \ll \mathcal{Y}_1 \sim 1$ fixed by the charm and top quark masses respectively, and with $\omega \sim 100$ TeV being an intermediate Left-Right symmetry breaking scale and $p \gg \omega$, associated to the breaking of the family symmetry. This results in the hierarchy $m_\text{B} \gg m_\text{S} \gg m_\text{D}$ with the latter expected to emerge not far above the TeV scale. As a particular benchmark scenario inspired by the phenomenologically relevant ranges of the SHUT model parameters found in Ref.~\cite{Morais:2020ypd}, we adopt the following values of the vector-like $D$-quark mixing elements in the extended CKM matrix:
\begin{equation}
V_{\rm uD} \simeq 5.1 \times 10^{-6} \,, \qquad
V_{\rm cD} \simeq 2.6 \times 10^{-5} \,, \qquad
V_{\rm tD} \simeq 0.016 \,.
\label{eq:CKM_matrix}
\end{equation}

In our numerical analysis below, we consider only light quark jets in the final states of VLQ decays $D\to W+$jets, so only $V_{\rm uD}$ and $V_{\rm cD}$ will be relevant in what follows. These will be fixed as above, whereas a numerical scan will be performed over the VLQ mass $m_{\rm D}$ within a phenomenologically acceptable range (see below).

%%%%%%%%%%%%%%%%%%%%%%%%%%%%%%%%%%%%%%
\section{VLQ search at the LHC}
\label{sec:Search}
%%%%%%%%%%%%%%%%%%%%%%%%%%%%%%%%%%%%%%

The generation of events in proton-proton collisions at the LHC follows the standard methods employed by the particle physics community. In this work, we use a specially designed chain of computational tools developed in the previous work of Ref.~\cite{Freitas:2020ttd}, and here we only briefly elaborate on the corresponding machinery. First, the Lagrangian density is constructed utilizing the \verb|SARAH| package \cite{Staub:2013tta}, which can be used to derive interaction vertices between the fields, their spectra and mixing. The latter are then outputted into \verb|UFO| \cite{Degrande:2011ua} files that can be used in Monte-Carlo event generators. In particular, here we employ \verb|MadGraph| \cite{Alwall:2014hca} for hard-scattering events and \verb|Pythia8| \cite{Sjostrand:2014zea} for hadronization/showering processes. Fast-simulation of the detector is implemented using \verb|Delphes| \cite{deFavereau:2013fsa} (with the ATLAS default card), while the kinematic data extraction is done in \verb|ROOT| \cite{Brun:1997pa}. We generate 250k events, for both VLQ production signal and backgrounds, at a centre-of-mass energy of $\sqrt{s}=14~\mathrm{TeV}$ and all at leading-order (LO) precision. We also employ the parton-distribution function NNPDF2.3 \cite{Ball:2013hta}, which fixes the strong-coupling constant, $\alpha_s$, and its evolution.

First, it is instructive to analyse the current constraints from direct searches of down-type VLQs. From the experimental side, among the well-known NP states a VLQ is one of the most searched for at the LHC. Since it is a coloured particle, its production rate is expected to be relatively high. In this regard, typical lower bounds on the VLQ mass range from 1.4 TeV up to 2.0 TeV (see the latest exclusion bounds, as of July 2021, in the summary plots of Ref.~\cite{ATLAS_twiki_VLQs}). However, one must carefully analyse the assumptions made in these analyses. First, the vast majority of current searches focus on dominant couplings to the third generation of chiral quarks (see e.g. Refs.~\cite{ATLAS:2018cjd,ATLAS:2018alq,ATLAS:2018uky,ATLAS:2017nap,ATLAS:2016scx,ATLAS:2015ktd}). Such an assumption is not the most general, as there is no reason for the VLQ to not couple to lighter quarks. Indeed, there have also been the searches assuming dominant light-quark couplings -- see, for instance, Refs.~\cite{ATLAS:2015lpr,ATLAS:2011tvb}. In Ref.~\cite{ATLAS:2015lpr}, in particular, a search for pair production of heavy VLQs has been performed, with their subsequent decays into light jets, that can go via either neutral ($D\rightarrow Zq$) or charged ($D\rightarrow Wq$) currents. This work excludes the VLQs with masses below 690 GeV at 95\% confidence level. Such a constraint is quite relevant since the same topology will be used in our analysis below. In the second work \cite{ATLAS:2011tvb}, the same VLQ decay modes into light jets have been considered, but in single VLQ production channels. For the neutral $D\rightarrow Zq$ decay channel, the VLQ mass has been constrained to be below 760 GeV, whereas a tighter lower bound of 900 GeV is obtained in the charged $D\rightarrow Wq$ decay channel. These constraints, however, are not relevant for the particular benchmark point that we have shown in the previous section. For example, we have calculated the production cross-section in \verb|MadGraph| for single production of the VLQ (using the benchmark point shown in Eq.~\eqref{eq:CKM_matrix}), for which we obtained $\sigma = 1.678 \times 10^{-4}$ pb for a mass of 800 GeV, which is consistent with the cross-section limits in Ref.~\cite{ATLAS:2011tvb}. For the sake of consistency, in this work we consider the VLQ mass to be greater than 800 GeV. In fact, it was verified that for the $D\rightarrow Wt$ channel this particular benchmark point yields a cross-section of $0.067~\mathrm{pb}$ compatible with the upper $95\%~\mathrm{C.L.}$ limit in \cite{ATLAS:2018ziw}, where $\sigma \approx 0.07~\mathrm{pb}$ and where it is assumed that the sum of the decay branching fractions to $Z b$ , $H b$ and $W t$ is one, as opposed to our scenario where $\mathrm{Br}(D \to W t) + \mathrm{Br}(D \to W q) \approx 1$.

\begin{figure}[ht!]
	\centering
    \includegraphics[width=0.40\textwidth]{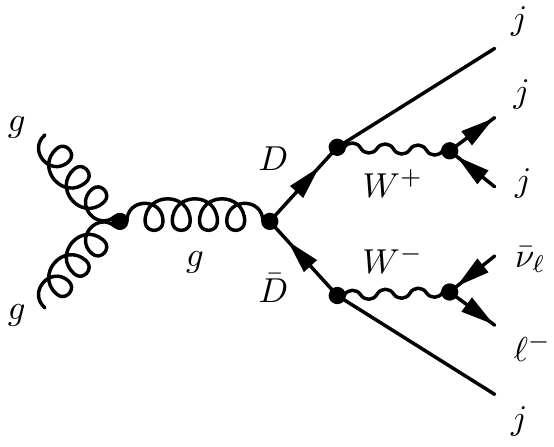}
	\caption{One of the LO Feynman diagrams for the pair-production of a $D$-type VLQ. Here, $g$ corresponds to gluons from initial colliding protons. The $D$-quark decays into a light jet $j = u, c,  \bar{u},\bar{c}$ and a $W$ boson, with one of the $W$'s further decaying via a leptonic channel, with one charged lepton ($\ell^- = e^-, \mu^-$) and the corresponding neutrino ($\bar{\nu}_e, \bar{\nu}_\mu$), and the other decaying via a hadronic channel into light jets. In the hadronic $W$ decay, both the up- and down-type SM quarks are considered.}
	\label{fig:VLQ-pair}
\end{figure}

As mentioned before, the chosen topology of the signal is equivalent to that of Ref.~\cite{ATLAS:2015lpr}, with VLQ pair-production undergoing via gluon fusion. An example for the LO Feynman diagram, with the contribution from the gluon triple vertex, is shown in Fig.~\ref{fig:VLQ-pair}. All tree-level graphs that contribute to the matrix-element are automatically included at the level of event generation in \texttt{MadGraph}. The signal is characterized by one charged lepton (either an electron or a muon) and missing energy, both arising due to $W^-$ boson decay. Additionally, four jets from first and second generation quarks are present, with two of these arising from the decay of one $W$ boson and the other two originating from the decays of the heavy VLQ states. Due to the presence of multiple jets, the matching and merging to the original partons is important. For this, we rely on the automatic matching procedure of \texttt{MadGraph}, which employs the kT-MLM scheme \cite{Mangano:2006rw}. For tagging the jets, we consider the following procedure. First, for each event, we begin by picking all jets and order them by its $p_T$ value. From this list, we subsequently impose the kinematic constraints, namely $|\eta(j)| < 2.5$ and $p_T(j)>25~\mathrm{GeV}$. From the surviving list we check whether the jets are tagged as coming from a $b$ quark. If the jet is not tagged as $b$, then it is accepted, else, it is discarded. If there are less than 4 jets that obey these constraints then the event is discarded. We do not consider any tau tagging information when selecting valid jet candidates.

The main irreducible backgrounds for this topology include:
\begin{itemize}
    \item Pair-production of top quarks, $t\bar{t}$, with one of the $W$ bosons decaying fully leptonically, whereas the other one decays into two light jets;
    \item Diboson production plus two light jets. In this case, one of the vector bosons decays into two light jets whereas the other one ($W$ boson) decays into a charged lepton and a neutrino;
    \item $V+\mathrm{jets}$, with $V$ either a $W$ or a $\mathrm{Z^0}$ boson. While for the former one considers the decay into a charged lepton and its corresponding neutrino, for the latter case we consider the fully leptonic decay channel (muon or electron modes).
\end{itemize}
Since the Monte-Carlo simulations are done at LO, we must reweight our events based on higher-order corrections available in literature. In particular, we consider approximate $\mathrm{N^3LO}$ corrections for $t\bar{t}$ \cite{Muselli:2015kba}, NLO corrections for $W+\mathrm{jets}$ \cite{Balossini:2009sa} and $WW+\mathrm{jets}$ \cite{Campbell:2011bn}, and NNLO corrections to $\mathrm{Z^0}+\mathrm{jets}$ \cite{Catani:2009sm}. 

While the NN models will do most of the heavy lifting, we will also consider simple selection criteria to help discriminate the signal events from the background. In particular, we consider that
\begin{equation}\nonumber
\begin{aligned}
&p_T(\ell^\pm) > 25 \text{ GeV} \,, \\
&\rm{MET}  > 20 \text{ GeV} \,, \\
&\abs{\eta(\ell^\pm)} \leq 2.5 \,,\\
\end{aligned}
\end{equation}
with $\eta = -\ln(\tan(\theta/2))$, $\theta$ is the angle with respect to the beam axis, and MET is defined as a missing transverse energy. With this, we extract all relevant kinematic and angular information about the final states' configurations, such as the transverse momentum, pseudo-rapidity and mass distributions of pairs of produced particles. Low-level observables include the transverse momenta, energy, pseudo-rapidity and the azimuthal angle of the four jets ($j_1$, $j_2$, $j_3$ and $j_4$) with the charged leptons, as well as MET. The jets are ordered with respect to their total transverse momenta, that is, $j_1$ is the leading jet with the highest $p_T$ and $j_4$ is that with the lowest $p_T$. Additionally, observables reconstructed from the combination of various objects are also considered. Namely, one uses the mass distributions $M(j_1,j_2)$, $M(j_1,j_3)$, $M(j_1,j_4)$, $M(j_2,j_3)$, $M(j_2,j_4)$, $M(j_3,j_4)$, $M (\mathrm{MET},\ell^-,j_1)$, $M (j_2,j_3,j_4)$, $M(\mathrm{MET},\ell^-,j_2)$, $M (j_1, j_3, j_4)$, the cosine of the polar angle $\cos \Delta \theta(\ell^-,\mathrm{MET})$, $\cos \Delta \theta(\ell^-,j_1)$, $\cos \Delta \theta(\ell^-,j_2)$, $\cos \Delta \theta(\ell^-,j_3)$, $\cos \Delta \theta(\ell^-,j_4)$, $\cos\Delta \theta(j_1,j_2)$, $\cos\Delta \theta(j_1,j_3)$, $\cos\Delta \theta(j_1,j_4)$, $\cos\Delta \theta (j_2,j_3)$, $\cos\Delta \theta(j_2,j_4)$, $\cos\Delta \theta(j_3,j_4)$, the azimuthal angle distributions $\Delta \Phi (j_1,\ell^-)$, $\Delta \Phi (j_2,\ell^-)$, $\Delta \Phi (j_3,\ell^-)$, $\Delta \Phi (j_4,\ell^-)$, $\Delta \Phi (j_1,j_2)$, $\Delta \Phi (j_1,j_3)$, $\Delta \Phi (j_1,j_4)$, $\Delta \Phi (j_2,j_3)$, $\Delta \Phi (j_2,j_4)$, $\Delta \Phi (j_3,j_4)$ and the $\Delta R$ distributions $\Delta R (j_1,\ell^-)$, $\Delta R (j_2,\ell^-)$, $\Delta R (j_3,\ell^-)$, $\Delta R (j_4,\ell^-)$, $\Delta R (j_1,j_2)$, $\Delta R (j_1,j_3)$, $\Delta R (j_1,j_4)$, $\Delta R (j_2,j_3)$, $\Delta R (j_2,j_4)$, $\Delta R (j_3,j_4)$\footnote{$\Delta R$ is defined as the Euclidean distance in the $(\eta,\phi)$ plane, i.e. $\Delta R(i,j)= \sqrt{\Delta\Phi(i,j)^2 + \Delta\eta^2(i,j)}$ with $\Delta\Phi(i,j) = \phi_j - \phi_i$ and $\Delta \eta(i,j) = \eta_j - \eta_i$. $\Delta \theta(i,j)$ is identically defined, with $\Delta \theta(i,j) = \theta_j - \theta_i$. Mass distributions are reconstructed from the observables inside parenthesis, e.g. $M(j_1, j_2) \equiv M(j^\mu_1 + j_2^\mu)$, where $j^\mu$ are four-momentum vectors.}.

%%%%%%%%%%%%%%%%%%%%%%%%%%%%%%%%%%%%%%%%%%%%%%%%
\subsection{LHC analysis: Hybrid Net approach}
\label{subsubsec:DL}
%%%%%%%%%%%%%%%%%%%%%%%%%%%%%%%%%%%%%%%%%%%%%%%%

In this section, we present the DL methodology used in our analysis, which involves a combination of both image and tabular datasets. The network, which we name as Hybrid Net in what follows, will take into account both data types in the classification of background and signal classes. The image datasets will be trained in a convolutional model, whereas kinematic data are trained through a linear sequential model, after which we combine the predictions of both.

\begin{figure}[htb!]
	\centering
	\subfloat[Signal event]{\includegraphics[width=0.30\textwidth]{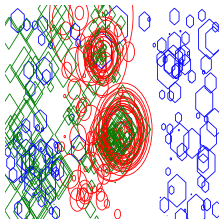}}
	\hspace{1.65em}
	\subfloat[$t\bar{t}$ event]{\includegraphics[width=0.30\textwidth]{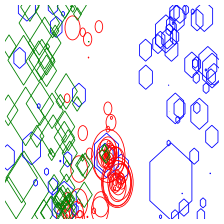}}
	\hspace{1.65em}
	\subfloat[$WW+$ jets event]{\includegraphics[width=0.30\textwidth]{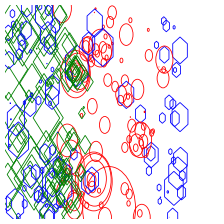}}
	\caption{Example of an abstract jet image in the $(\eta,\phi)$ plane. Green polygons are representative of photons, red circles are charged particles while blue hexagons indicate neutral hadrons. The radius of the polygons is proportional to the energy left by the particle in the calorimeter tower. In this image we have an event for a proton-proton collision at $\sqrt{s}=14$ TeV, with subsequent VLQ double production (to the right) and the corresponding background events ($t\bar{t}$ in the centre and $WW+$ jets in the right).}
	\label{fig:jet_images}
\end{figure}

Jet images, first introduced in Refs.~\cite{Cogan:2014oua,deOliveira:2015xxd}, are constructed by mapping particle hits in the detector into a grid in the $(\eta,\phi)$ plane, where the intensity of the pixel is directly proportional to the energy deposited by the particles. These methods, however, can be quite inefficient since in a typical jet-image there can exist plenty of blank pixels, making it computationally inefficient when passing through convolutional layers in the training phase. While various approaches have been proposed in the literature (such as particle clouds \cite{Qu:2019gqs} or through the use of graph networks \cite{Ju:2020xty,Farrell:2018cjr}), in this work we consider abstract jet-images as first discussed by one of the authors and collaborators in Ref.~\cite{Alves:2019ppy}. In this approach, besides associating each pixel with the energy of a particle, we also map characteristics of the particle into various geometrical shapes. For example, if it is a charged particle in a detector, we map it to a circle. In this scenario, the radius of the polygons is directly proportional to the particle energy. With this, we then provide additional features for the neural model to learn. Examples of these types of images can be seen in Fig.~\ref{fig:jet_images}. 
%%%%%%%%%%%%%%%%%%%%%%%%%%%%%%%%%%%%%%%%
%%%%%%%%%%%%%%%%%%%%%%%%%%%%%%%%%%%%%%%%

In order to construct such images, we impose the following conditions:
\begin{itemize}
    \item Charged particles hitting the detector are mapped into blue circles, whose radius is given as $r = \log{p_T}$, where $p_T$ is the transverse momentum of the particle;
    \item Neutral hadrons are mapped into hexagons, with the radius given as $r = \log(E_T)$, where $E_T$ is the transverse energy;
    \item Photons are mapped into squares, with the radius given by the same formula as the neutral hadrons.
\end{itemize}
The resulting images are then stored in the \verb|.png| format, with the pixel size of $224\times224$, requiring that both hadrons and photons are restricted to be in the kinematic ranges of $\abs{\eta} \leq 5.0$ and $\abs{\phi} \leq \pi$.

We use both the image and tabular information and feed it into a NN. The images pass through a convolutional network, whereas the corresponding kinematic information is passed through a linear model. We then combine the results by summing together the predictions of both networks leading to the final result. The hyper-parameter optimization designed to maximize the statistical significance is implemented via evolutionary algorithms as described in previous works \cite{Morais:2021ead,Bonilla:2021ize,Freitas:2020ttd}. Training of the NNs based on the statistical significance follows the analysis first performed by A. Elwood and D. Kr{\"u}cker \cite{Elwood:2018qsr}. In this work, only linear networks are part of the optimization procedure, whereas the convolutional network for the images is fixed to a ResNet-34 network \cite{he2015deep}. Furthermore, we consider the following list of hyper-parameters:
\begin{itemize}
    \item Number of neurons of the sequential model: 256, 512, 1024 and 2048;
    \item Number of layers of the sequential model: 1, 2, 3, 4 and 5;
    \item Bottleneck architecture\footnote{We define a bottleneck architecture as a neural model with an ever decreasing number of neurons at each layer.}: True or False;
    \item Initialisers in the sequential model: Normal, He normal and He uniform;
    \item Optimisers in the sequential model: Adam, SGD, Adamax and nadam;
    \item Vision models: ResNet-34.
\end{itemize}
The evolutionary algorithm builds sets of Hybrid Net models. Both the linear and convolution parts of the algorithms are constructed using \verb|PyTorch| \cite{paszke2019pytorch}. In the convolution part, we also utilized \verb|FastAI| \cite{2020arXiv200204688H}, which employs \verb|PyTorch| as backend. Given this methodology, let us now turn to a discussion of numerical results.

%%%%%%%%%%%%%%%%%%%%%%%%%%%%%%%%%%%%
\subsection{Numerical results}
\label{subsubsec:results}
%%%%%%%%%%%%%%%%%%%%%%%%%%%%%%%%%%%%

\begin{figure}[htb!]
	\centering
	\includegraphics[width=\textwidth]{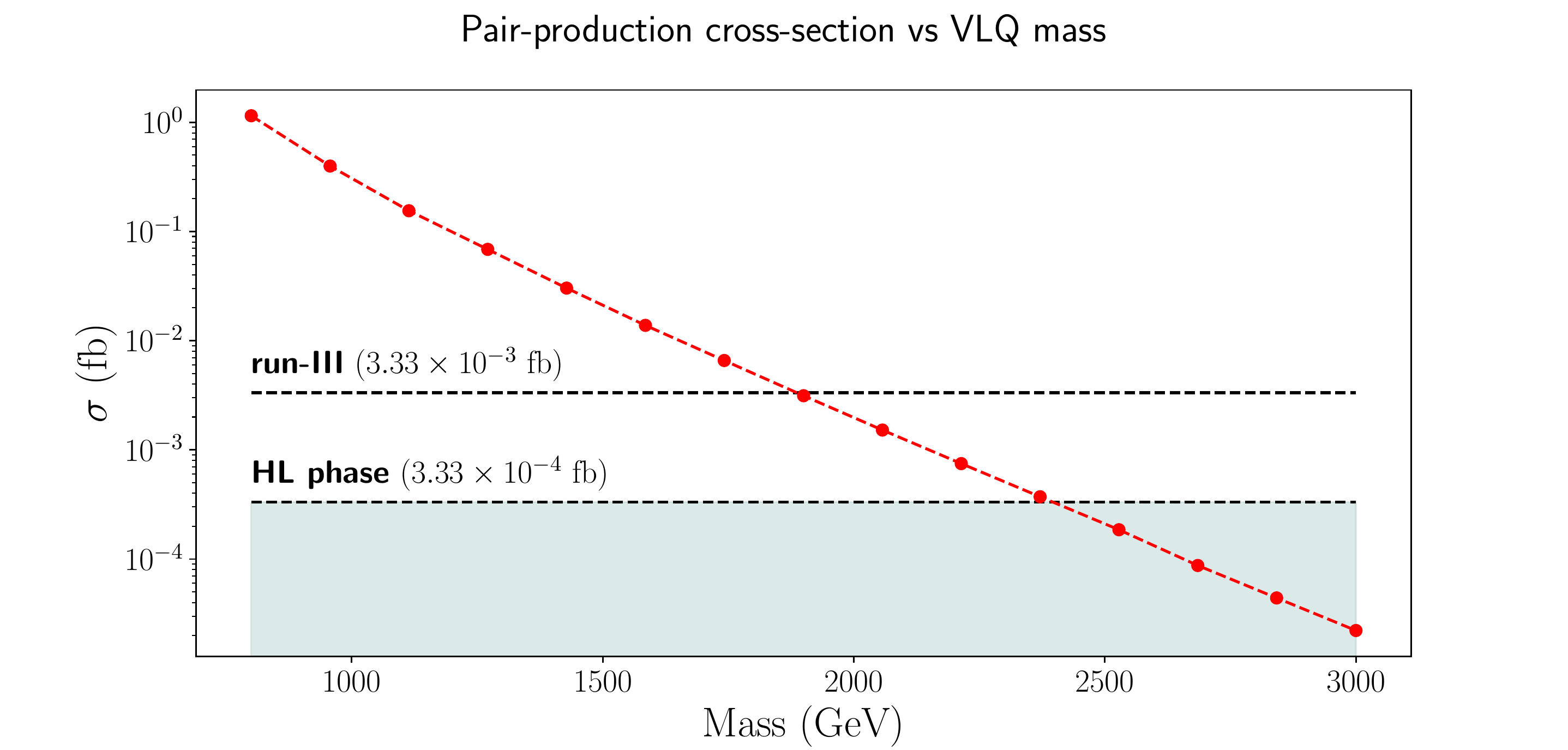}
	\caption{The production cross-section (in femtobarn) as a function of the VLQ mass (in GeV). Two horizontal dashed lines are shown, one for the run-III target luminosity and the other for the HL phase of the LHC, which represent the minimal cross section for which a single event can be produced, i.e. $N_{\mathrm{exp}}=\sigma\mathcal{L} = 1$ for the dashed lines. The area filled in gray represents the region where neither the HL-LHC or run-III LHC have the required sensitivity to detect the $D$-VLQ for the chosen benchmark~\eqref{eq:CKM_matrix}. The $y$-axis is shown in logarithmic scale.}
	\label{fig:xsec_points}
\end{figure}

We first begin by presenting the results for a particular (fixed) VLQ mass. Focusing on the lightest $D$-VLQ with a mass of $800~\mathrm{GeV}$ and applying the selection criteria described above, the obtained LO cross-sections for the signal, and corresponding higher-order rates for the backgrounds, are showcased in  Tab.~\ref{tab:cross_sections}. Additionally, the total expected number of events at both run-III and the HL phase of the LHC are also indicated. The variation of the post-selection cross-section with the VLQ mass is shown in Fig.~\ref{fig:xsec_points}. In this figure, we also plot dashed lines to indicate the allowed sensitivity expected for future runs of the LHC. In particular, we show that for the benchmark values of the VLQ-light quark mixing elements indicated in Eq.~\eqref{eq:CKM_matrix}, the run-III phase of the LHC can potentially probe $D$-type VLQs up to 1.9 TeV in this channel, while the HL phase has a sufficient sensitivity to exclude VLQs up to 2.37 TeV. Note that this approximate limits are set for the best case scenario, that is, in the limit of experimental errors going to zero and therefore can be seen as guideline to the mass scale where the VLQ can be potentially probed. Indeed, in the scenario where only one event is produced, the systematics of the backgrounds completely overshadow the signal.

\begin{table}[htb!]
\centering
\captionsetup{justification=raggedright,singlelinecheck=false}
	\resizebox{1.0\textwidth}{!}{\begin{tabular}{c|c|c|c|c}
			& $\sigma$ (before cuts, in fb) & $\sigma$ (after cuts, in fb) & Events at run-III & Events at HL-LHC \\ \hline
			VLQ 800 GeV &	
			\makecell{
				\begin{math}
				1.402
				\end{math}
			}
			&
			1.063
			&
			318
			&
			3189
		    \\
		    \hline
		    $\mathrm{WW}+\mathrm{jets}$ &
			\makecell{
			\begin{math}
			1.24\times 10^{5}
			\end{math}
			}
			&
			\begin{math}
			7.72 \times 10^{3}
			\end{math}
			&
			\begin{math}
			2.316\times 10^6
			\end{math}
			&
			\begin{math}
			2.316\times 10^7
			\end{math}
			\\
			\hline 
			$\mathrm{W}+\mathrm{jets}$ &
			\makecell{
			\begin{math}
			7.95\times 10^{8}
			\end{math}
			}
			&
			\begin{math}
			3.46 \times 10^{5}
			\end{math}
			&
		    \makecell{
			\begin{math}
			1.038\times 10^8
			\end{math}
			}
			&
		    \makecell{
			\begin{math}
			1.038\times 10^9
			\end{math}
			}
			\\
			\hline      
			$\mathrm{Z^0}+\mathrm{jets}$  &
    		\makecell{
			\begin{math}
			6.33\times 10^{7}
			\end{math}
			}
			&
			\begin{math}
			5.70\times 10^{3}
			\end{math}
			&
			\begin{math}
			1.71\times 10^6
			\end{math}
			&
			\begin{math}
            1.71\times 10^7
            \end{math}
			\\
			\hline
			$t\bar{t}$  &
			\makecell{
			\begin{math}
			9.89\times 10^{5}
			\end{math}
			}
			&
			\begin{math}
			1.83\times 10^{5}
			\end{math}
			&
			\makecell{
			\begin{math}
			5.49\times 10^7
			\end{math}
			}
			&
			\makecell{
			\begin{math}
			5.49\times 10^8
			\end{math}
			}
			\\
			\hline
	\end{tabular}}
	\captionsetup{justification=raggedright,singlelinecheck=false}
	\caption{Predicted total cross section (in fb) for both the signal and each respective background, before and after the cut selection. In the last two columns we indicate the overall number of expected events for the analysis, calculated as $N=\sigma \mathcal{L}$, for run-III and the HL phase of the LHC.}
	\label{tab:cross_sections}
\end{table}

\begin{figure}[]
	\centering
	\includegraphics[width=\textwidth]{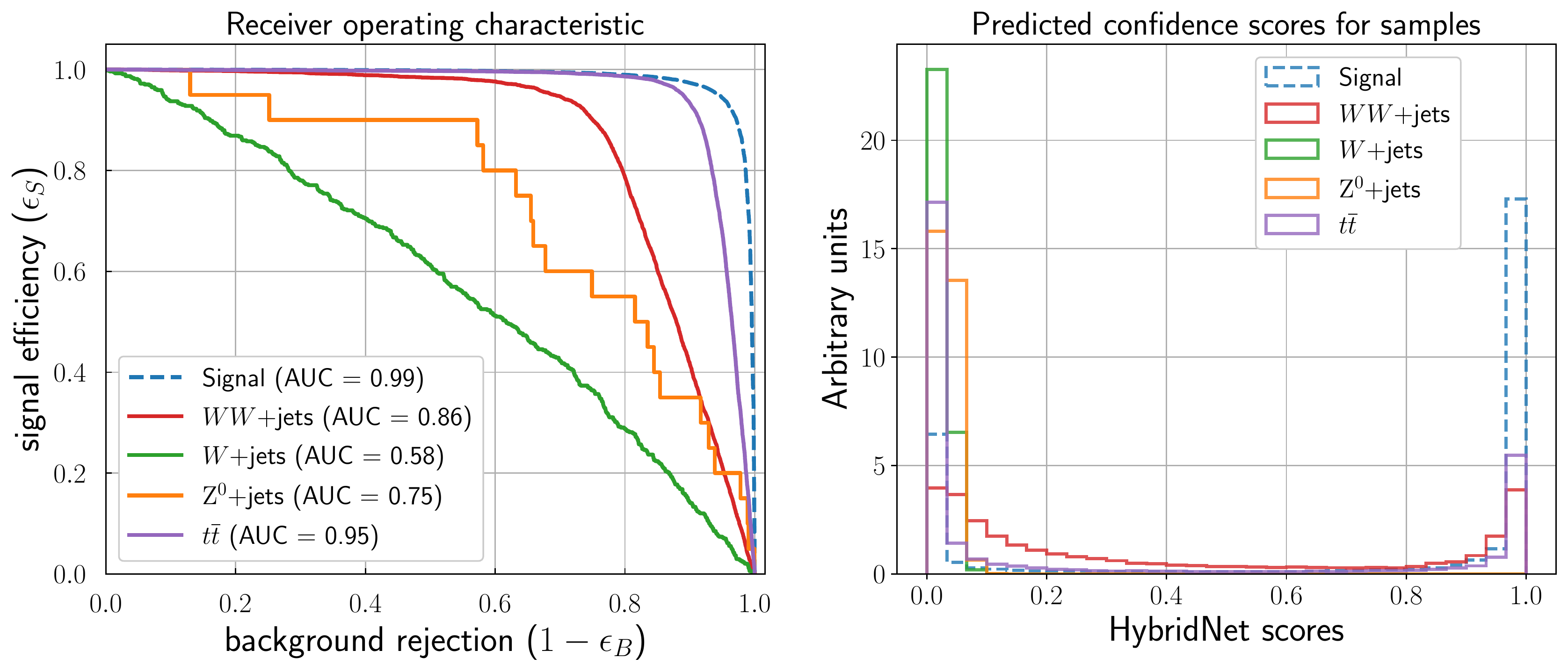} \\
	\includegraphics[width=\textwidth]{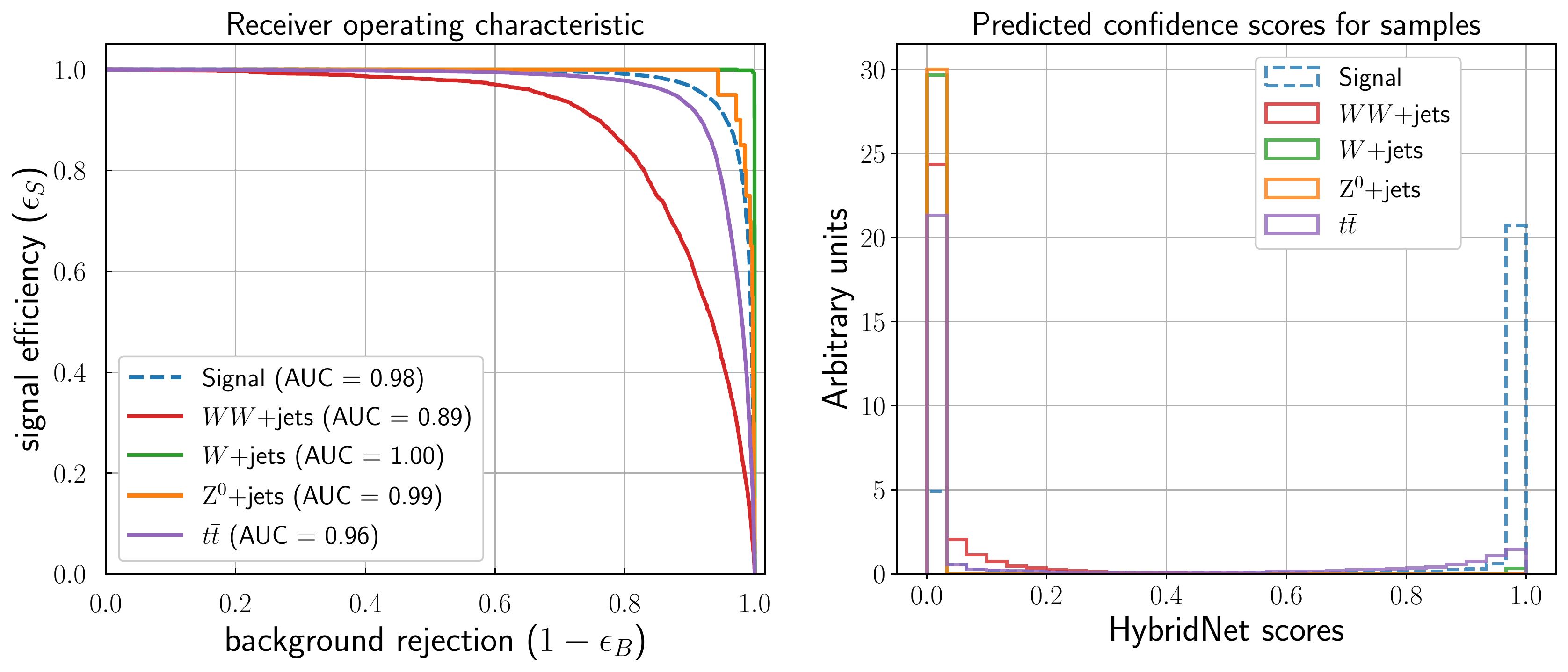} \\
	\includegraphics[width=\textwidth]{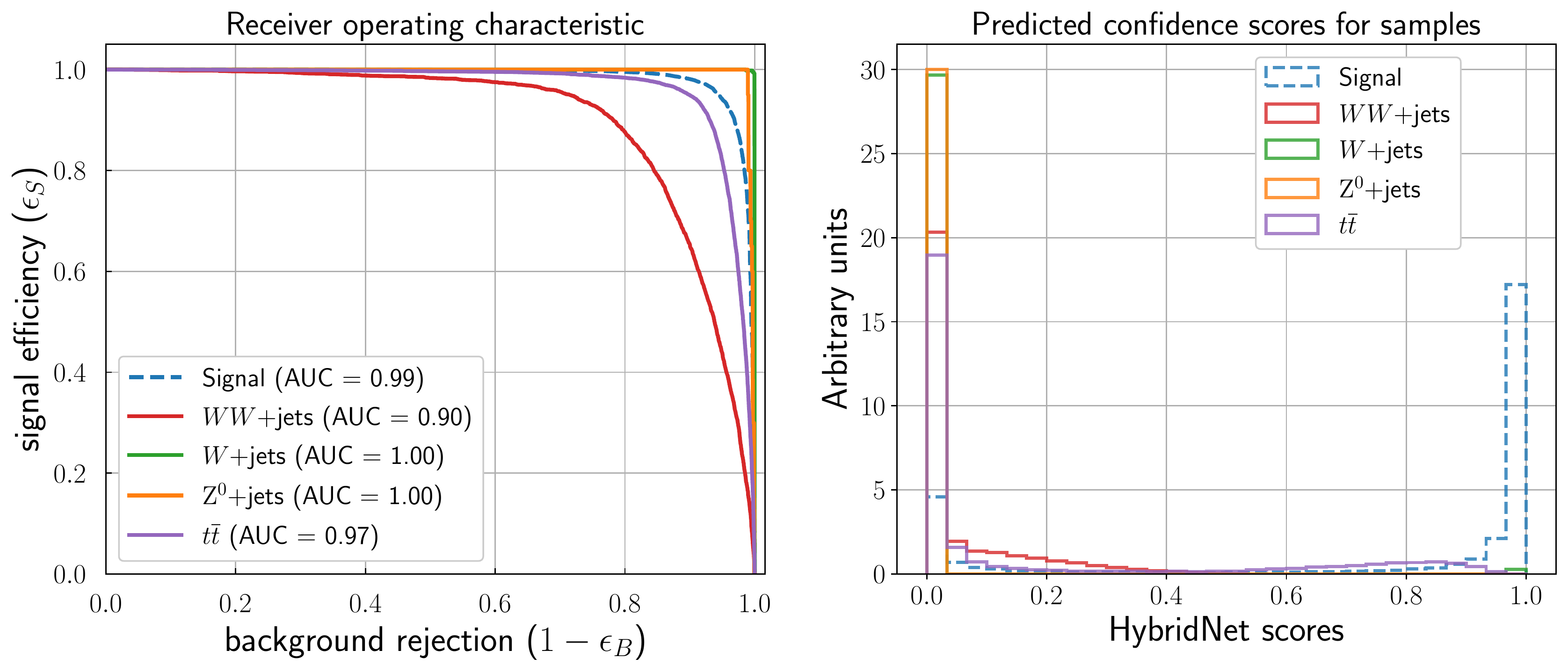}
	\caption{Performance results for the Hybrid network. On each panel, we plot the ROC on the left-hand side and, on the right, -- the prediced confidence score for our samples. In the top panel, the results are shown for classification with only kinematic data, in the middle panel we have the results only for classification with the jet images, and at the bottom, the results when both data types were part of the training. The signal is indicated with a dashed blue line, while the $WW$+jets background -- in red, the $W$+jets -- in green, $\mathrm{Z^0}$+jets -- in orange and $t\bar{t}$ -- in purple.}
	\label{fig:ROC_and_PCS}
\end{figure}

In order to evaluate the performance of the network, we show in Fig.~\ref{fig:ROC_and_PCS} the receiver operating characteristic (ROC) plots for three distinct situations: for classification with only kinematic data (top panel), for classification with only jet images (middle panel) and classification where both are taken into account (bottom panel). These plots can serve as a measurement of how well the network performs on the data. In particular, these results are shown for the validation dataset, i.e.~for the dataset that the network was not trained on. Within the labeling of the curves, we also indicate the area under the curve (AUC), which is related to the accuracy. As one can readily observe, the combination of both the jet images and the kinematic data offers the greater discriminating power with the highest accuracy, clearly validating the improved performance of the developed hybrid network. Indeed, we have found that the tabular data is one that performs the worst, with the jet image classification giving the best results. Such a good accuracy indicates an efficient separation between the signal and background classes. Further confirmation of this fact can be given by the predicted confidence scores for the samples plotted on the left-hand-side of Fig.~\ref{fig:ROC_and_PCS}. Here, we can clearly see that the model successfully separates the signal classes (mainly populating the convolutional NN scores of around one) from the background (which mainly populates a score of around zero). Of course, such separation is not perfect, with some contamination still present, in a sense that some background and signal overlap in some regions of the convolutional NN parameter domain. All subsequent analysis on the data will be done based on the best results, i.e.~on the hybrid network scores with both kinematic and image data.
%%%%%%%%%%%%%%%%%%%%%%%%%%%%%%%%%%%%%%%%%%%%%%%%%%%
%%%%%%%%%%%%%%%%%%%%%%%%%%%%%%%%%%%%%%%%%%%%%%%%%%%

The histograms with the kinematic and angular observables that are utilized by the NN are plotted in appendix, see Figs.~\ref{fig:Dimensionfull-vars} and \ref{fig:Dimensionless-vars}. One can observe that the angular distributions for the signal tends to be overshadowed by the backgrounds, with the signal following the same structure as the background, with the only exception being the pseudo-rapidity distributions. Note that the signal has a dominant peak at $\eta\sim 0$, whereas the backgrounds spread over the entirety of the allowed range $\abs{\eta}\leq 2.5$. The kinematics, on the other hand, offers the most significant discriminating power, with distributions being peaked in kinematic regions where the SM backgrounds are less dominant. This is true for all transverse momentum, energy and MET distributions. Similarly, mass distributions also peak in higher invariant mass domains compared to those of the SM backgrounds.

Each individual event corresponds to a particular value of the physical observables transported over into the tabular data. Each event will have an associated jet image. A combination of both is used in the classification procedure, and then the statistical significance is computed using the validation dataset. We present the results using the Asimov estimate, which is defined as \cite{Cowan:2010js}
\begin{equation}\label{eq:Asimov_sig}
\mathcal{Z}_A = \Bigg[2\Bigg((s + b)\ln\Bigg(\frac{(s+b)(b+\sigma_b^2)}{b^2 + (s+b)\sigma_b^2}\Bigg) -\frac{b^2}{\sigma_b^2}\ln\Bigg(1+\frac{\sigma_b^2 s}{b(b+\sigma_b^2)}\Bigg)\Bigg)\Bigg]^{1/2} \,,
\end{equation}
where $s$ is the number of signal events, $b$ is the number of background events and $\sigma_b$ is the uncertainty of the background. For this metric, we consider three scenarios with either $\sigma_b = 15\%$, $\sigma_b = 25\%$ or the much more conservative scenario of $\sigma_b = 50\%$ systematic uncertainty. The first two are the most realistic. 

%%%%%%%%%%%%%%%%%%%%%%%%%%%%%%%%%%%%%%%%%%%%%%%%%%%
%%%%%%%%%%%%%%%%%%%%%%%%%%%%%%%%%%%%%%%%%%%%%%%%%%%%
%%%%%%%%%%%%%%%%%%%%%%%%%%%%%%%%%%%%%%%%%%%%%%%%%%%%
\begin{figure*}[t!]
    \captionsetup{justification=raggedright,singlelinecheck=false}
	\includegraphics[width=\textwidth]{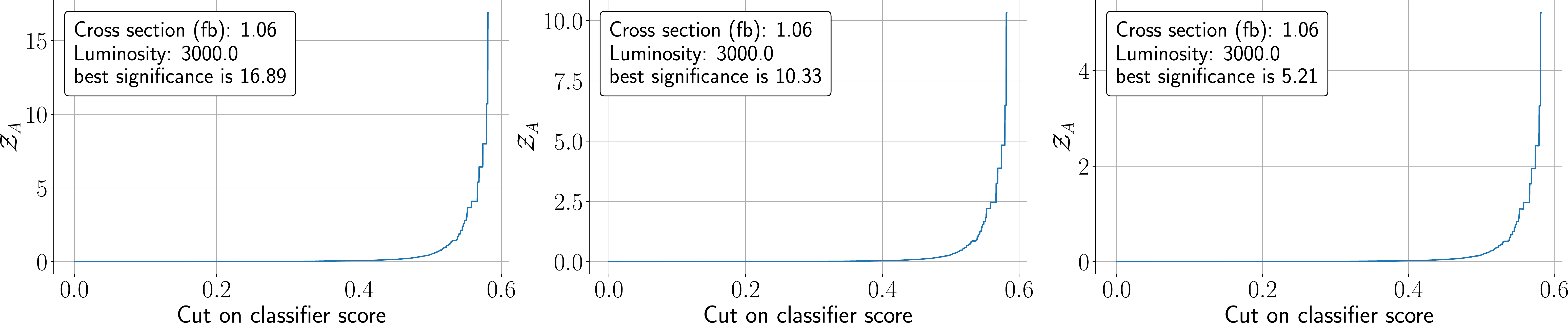}\\
	\caption{The statistical significance for a hypothetical discovery of VLQs as a function of the NN score for different values of the systematic uncertainties considered in this work. The VLQ mass is 800 GeV, and the collider luminosity is $\mathcal{L}=3000~\mathrm{fb}$. The significance is computed following the implementation of an evolution algorithm whose method is described in \cite{Freitas:2020ttd}. From left to right, we have $\sigma_b = 15\%$, $\sigma_b = 25\%$ and $\sigma_b = 50\%$. 
		\label{fig:sig_plots}}
\end{figure*}
%%%%%%%%%%%%%%%%%%%%%%%%%%%%%%%%%%%%%%%%%%%%%%%%%%%%
%%%%%%%%%%%%%%%%%%%%%%%%%%%%%%%%%%%%%%%%%%%%%%%%%%%%
\begin{figure}[hb!]
	\centering
	\includegraphics[width=\textwidth]{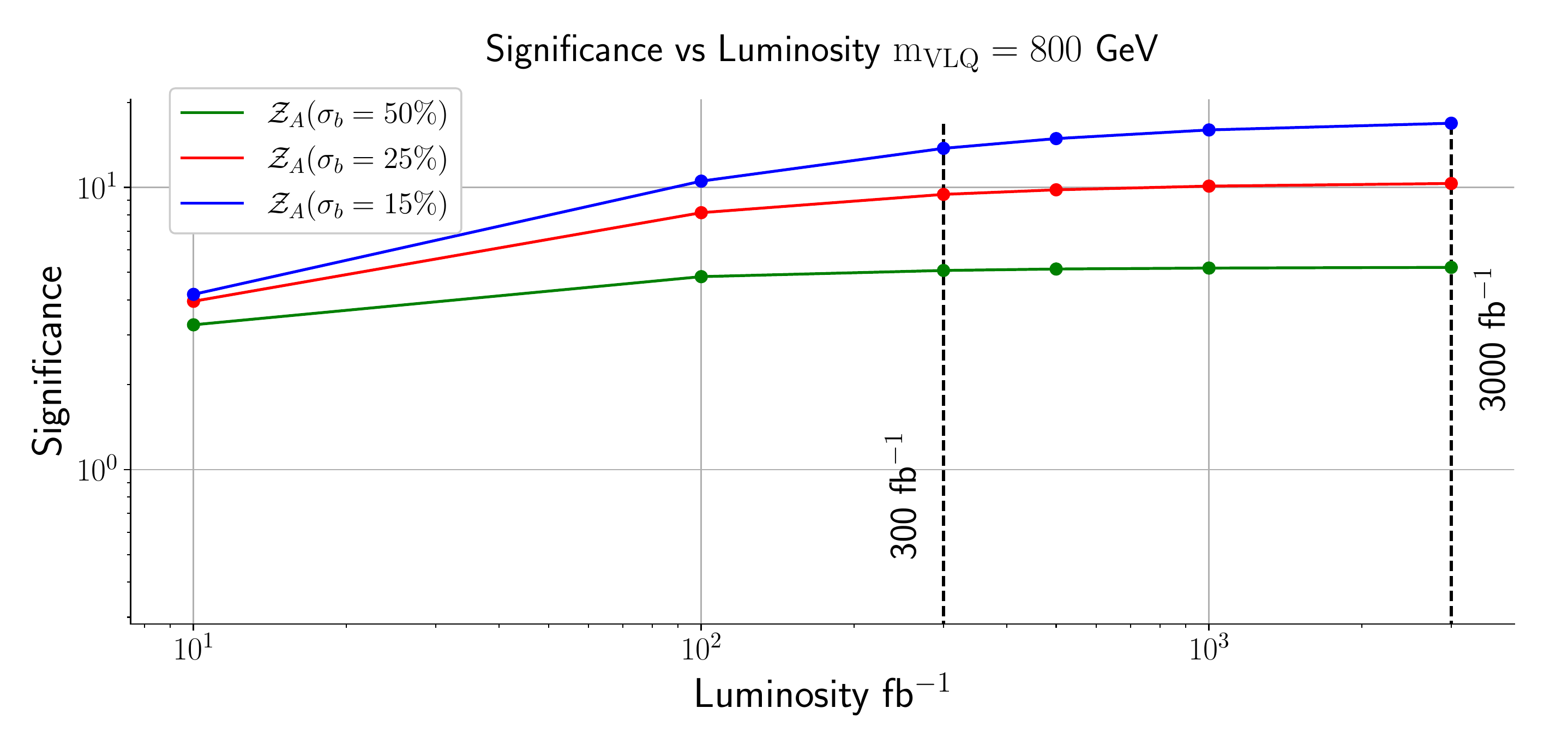}
	\caption{ The statistical significance as a function of the systematic uncertainties for pair-production of 800 GeV $D$-VLQs. The green curve corresponds to $\sigma_b = 50\%$, the red curve -- to $\sigma_b=25\%$ and the blue line -- to $\sigma_b = 15\%$. There are two vertical dashed lines marking the run-III luminosity ($300~\mathrm{fb^{-1}}$) and the HL-LHC luminosity ($3000~\mathrm{fb^{-1}}$). Both the $y$-axis and $x$-axis are shown in logarithmic scale.}
	\label{fig:lum_vs_sig}
\end{figure}
%%%%%%%%%%%%%%%%%%%%%%%%%%%%%%%%%%%%%%%%%%%%%%%%%%%%
Considering the HL phase of the LHC, which is expected to reach an integrated luminosity of $\mathcal{L} = 3000~\mathrm{fb^{-1}}$, we can determine the median significance for the considered metrics as a function of the NN score, as shown in Fig.~\ref{fig:sig_plots}. The maximal significances for each metric are then determined to be
\begin{itemize}
    \item $\mathcal{Z}_A(\sigma_b = 15\%) = 16.89\sigma$,
    \item $\mathcal{Z}_A(\sigma_b = 25\%) = 10.33\sigma$,
    \item $\mathcal{Z}_A(\sigma_b = 50\%) = 5.21\sigma$,
\end{itemize}i.e.~we obtain significances greater than $5\sigma$ for all systematic uncertainties, even in the most conservative scenario of $50\%$. Such a strong signature of VLQs at the HL run of the LHC can be expected, since VLQs are coloured particles, and their production is enhanced at hadron colliders. 

It is also relevant to analyse the impact for other luminosities, in particular, for the run-III of the LHC, which is set to deliver $\mathcal{L}=300~\mathrm{fb^{-1}}$ of data. In Fig.~\ref{fig:lum_vs_sig} we plot the statistical significance as a function of the systematic errors. Specifically, the maximal significances obtained for the run-III luminosity read as follows:
\begin{itemize}
    \item $\mathcal{Z}_A(\sigma_b = 15\%) = 13.76\sigma$,
    \item $\mathcal{Z}_A(\sigma_b = 25\%) = 9.45\sigma$,
    \item $\mathcal{Z}_A(\sigma_b = 50\%) = 5.08\sigma$,
\end{itemize}
i.e. the significance is above $5\sigma$ for all considered systematics, indicating that the VLQ production can also be probed at run-III.
%%%%%%%%%%%%%%%%%%%%%%%%%%%%%%%%%%%%%%%%%%%%%%%%%%%
\begin{table}[htb!]
\begin{center}
\captionsetup{justification=raggedright,singlelinecheck=true}
\begin{tabular}{c|c|l|l|cll|c|l|l|c|l|l}
Benchmarks           & \multicolumn{3}{c|}{$800~\mathrm{GeV}$}                    & \multicolumn{3}{c|}{$1200~\mathrm{GeV}$}                    & \multicolumn{3}{c|}{$1700~\mathrm{GeV}$} & \multicolumn{3}{c|}{$2200~\mathrm{GeV}$}              \\[2mm] \hline
\multirow{3}{*}{\makecell{$\hphantom{.}$ \quad \hspace{2.6em}$\mathcal{Z}_A(\sigma_b = 15\%)$ \\ \hspace{-0.7em} $300~\mathrm{fb^{-1}}$ \hspace{0.3em}$\mathcal{Z}_A(\sigma_b = 25\%)$ \\ \hspace{3.9em} $\mathcal{Z}_A(\sigma_b = 50\%)$}} & \multicolumn{3}{c|}{\redBU $13.76\sigma$\Tstrut}      & \multicolumn{3}{c|}{$1.23\sigma$}      & \multicolumn{3}{c|}{$0.06\sigma$} & \multicolumn{3}{c|}{$-$}           \\ \cline{2-13}
                     & \multicolumn{3}{c|}{\redBU $9.45\sigma$}     & \multicolumn{3}{c|}{$0.89\sigma$}     & \multicolumn{3}{c|}{$0.04\sigma$}   & \multicolumn{3}{c|}{$-$}\\ \cline{2-13}
                     & \multicolumn{3}{c|}{\redBU $5.08\sigma$\Bstrut} & \multicolumn{3}{c|}{$0.49\sigma$} & \multicolumn{3}{c|}{$0.02\sigma$} & \multicolumn{3}{c|}{$-$}  \\ \hhline{=============}
\multirow{3}{*}{\makecell{$\hphantom{.}$ \quad \hspace{2.6em}$\mathcal{Z}_A(\sigma_b = 15\%)$ \\ \hspace{-0.9em} $1000~\mathrm{fb^{-1}}$ \hspace{0.3em}$\mathcal{Z}_A(\sigma_b = 25\%)$ \\ \hspace{3.9em} $\mathcal{Z}_A(\sigma_b = 50\%)$}} & \multicolumn{3}{c|}{\redBU $15.99\sigma$\Tstrut}      & \multicolumn{3}{c|}{$1.51\sigma$}      & \multicolumn{3}{c|}{$0.07\sigma$} & \multicolumn{3}{c|}{$-$}       \\ \cline{2-13}
                     & \multicolumn{3}{c|}{\redBU $10.11\sigma$}     & \multicolumn{3}{c|}{$0.98\sigma$}     & \multicolumn{3}{c|}{$0.05\sigma$} & \multicolumn{3}{c|}{$-$}     \\ \cline{2-13}
                     & \multicolumn{3}{c|}{\redBU $5.18\sigma$\Bstrut} & \multicolumn{3}{c|}{$0.51\sigma$} & \multicolumn{3}{c|}{$0.02\sigma$} & \multicolumn{3}{c|}{$-$} \\ \hhline{=============}
\multirow{3}{*}{\makecell{$\hphantom{.}$ \quad \hspace{2.6em}$\mathcal{Z}_A(\sigma_b = 15)\%$ \\ \hspace{-0.9em} $3000~\mathrm{fb^{-1}}$ \hspace{0.3em}$\mathcal{Z}_A(\sigma_b = 25)\%$ \\ \hspace{3.9em} $\mathcal{Z}_A(\sigma_b = 50)\%$}} & \multicolumn{3}{c|}{\redBU $16.89\sigma$\Tstrut}      & \multicolumn{3}{c|}{$1.64\sigma$}      & \multicolumn{3}{c|}{$0.08\sigma$} & \multicolumn{3}{c|}{$0.014\sigma$}          \\ \cline{2-13}
                     & \multicolumn{3}{c|}{\redBU $10.33\sigma$}     & \multicolumn{3}{c|}{$1.01\sigma$}     & \multicolumn{3}{c|}{$0.05\sigma$} & \multicolumn{3}{c|}{$0.0071\sigma$}   \\ \cline{2-13}
                     & \multicolumn{3}{c|}{\redBU $5.21\sigma$\Bstrut} & \multicolumn{3}{c|}{$0.51\sigma$} & \multicolumn{3}{c|}{$0.02\sigma$} & \multicolumn{3}{c|}{$0.0036\sigma$}  \\ \hhline{=============}
\end{tabular}
\caption{The statistical significance for different values of the systematics considered in this work, for various masses of the VLQ and different collider luminosities. Scenarios which pass the $5\sigma$ requirement for potential exclusion (or hypothetical discovery) are displayed in {\redBU red}. Points marked with ``--'' indicate that the cross-section is not enough to produce one event at the corresponding luminosity.}
\label{tab:VLQ_significance_table}
\end{center}
\end{table}
%%%%%%%%%%%%%%%%%%%%%%%%%%%%%%%%%%%%%%%%%%%%%%%%%%%

In order to complement these results further, we implement the same machinery in a numerical scan over the VLQ mass, up to the detectability limit of the HL phase, which corresponds up to the VLQ mass of around 2.2 TeV. For these additional mass points, we have used the same network as the one for the 800 GeV benchmark, with the results summarised in Table~\ref{tab:VLQ_significance_table}. As the main result, a 5$\sigma$ significance is obtained for VLQ masses of up to 800 GeV. In particular, for a VLQ of 1.2~TeV, the highest significance was of 1.64$\sigma$, at the HL phase of the LHC. Indeed, for points with a larger mass, a significance above the discovery threshold requires a greater control of the background systematics. For example, if we consider a systematic uncertainty at the percent level at the high-luminosity phase of the LHC, for masses of order $1.2~\mathrm{TeV}$ one obtains a remarkable enhancement in the significance up to $\mathcal{Z}_A = 17.82\sigma$, while for a $1.7~\mathrm{TeV}$ VLQ we could expect up to $\mathcal{Z}_A = 3.68\sigma$. These results demonstrate that for larger masses systematic uncertainties become increasingly relevant and, in the best interest of the LHC physics programme, must be well understood and studied by our experimentalist colleagues. In principle, it is expected that an improvement in the systematic uncertainties is achieved over time opening up the possibility to probe larger VLQ masses above 800 GeV in the context of the scenario discussed in this article. Note that this picture can be potentially improved through the combination of various production channels, namely single-production and double production channels, such as the ones involving decays into to top quarks, which may also be relevant for the considered model.

%%%%%%%%%%%%%%%%%%%%%%%%%%%%%%%%%%%%%%%%%
\section{Conclusions}
\label{sec:conclusions}
%%%%%%%%%%%%%%%%%%%%%%%%%%%%%%%%%%%%%%%%%

In this paper, we continue the DL phenomenological analysis, as first started in our previous work \cite{Freitas:2020ttd}, of exotic vector-like fermions that naturally arise in the low-energy limit of a F-GUT model based on the trinification gauge group. In particular, we have focused on possible collider signatures of the weak-isospin-singlet $D$-type VLQ at the LHC, via pair-production topologies, with subsequent VLQ decay into lighter chiral quarks. The latter topology results in a single charged lepton, light jets and a neutrino in the final state particularly convenient for experimental analysis at the LHC. In order to accomplish this task, we have performed Monte-Carlo simulations and utilized sophisticated DL methods for efficient separation of SM backgrounds from the VLQ signal. This enables us to compute the statistical significance for a hypothetical discovery of the considered VLQ. In the context of the VLQ detection prospects at the LHC, we have utilised an evolutionary algorithm that optimises the construction of a NN based on the maximisation of the statistical significance. For this purpose, we have exploited both the kinematic tabular data as well as the detector jet images as inputs in the classification procedure through the use of a Hybrid Net. We have found that the combination of both data types improves the accuracy with respect to their individual contribution.

With this technique, we have found that the down-type VLQs with a mass of up to 800 GeV can be excluded both at the run-III and the HL phase of the LHC in the decay channels to light quarks. In particular, we have determined that at $\mathcal{L}=3000~\mathrm{fb^{-1}}$ one rules out VLQ at this mass scale with statistical significance of $\mathcal{Z}_A = 16.89\sigma$, assuming systematic uncertainties of $\sigma_b = 15\%$. For this same point, we have found $\mathcal{Z}_A = 13.76\sigma$ for run-III luminosity. Additionally, we have found that the statistical significance of heavier VLQs decaying to light quarks remains bellow the discovery threshold and an exclusion bound can not be promptly determined for the considered channel. Instead, it was shown that a better control over systematic uncertainties is required in order to open up the possibility to probe larger masses in the light quarks channel.

%%%%%%%%%%%%%%%%%%%%%%%%%%%%
\section*{Acknowledgments}
%%%%%%%%%%%%%%%%%%%%%%%%%%%%

The work developed in this article is supported by the projects PTDC/FIS-PAR/31000/2017 and CERN/FIS-PAR/0014/2019. J.G., F.F.F., and A.P.M. are supported by the Center for Research and Development in Mathematics and Applications (CIDMA) through the Portuguese Foundation for Science and Technology (FCT - Funda\c{c}\~{a}o para a Ci\^{e}ncia e a Tecnologia), references UIDB/04106/2020 and UIDP/04106/2020.
A.P.M.~is also supported by national funds (OE), through FCT, I.P., in the scope of the framework contract foreseen in the numbers 4, 5 and 6 of the article 23, of the Decree-Law 57/2016, of August 29, changed by Law 57/2017, of July 19.
J.G. is also directly funded by FCT through a doctoral program grant with the reference 2021.04527.BD.
R.P.~is supported in part by the Swedish Research Council grant, contract number 2016-05996, as well as by the European Research Council (ERC) under the European Union's Horizon 2020 research and innovation programme (grant agreement No 668679).

\appendix

%%%%%%%%%%%%%%%%%%%%%%%%%%%%%%%%
\section{Low-scale LS-SHUT Model}\label{app:Model}
	Here, we consider a possible low-energy realisation of the Yukawa sector of the LS-SHUT model. The full physical Lagrangian %density 
	of the model was previously %already 
	shown in Ref.~\cite{Freitas:2020ttd}. %as such, 
	Here, we are %only 
	going to specify %showcase 
	the relevant terms that impact the (chiral and vector-like) quark sector. %All 
	The charges of %quantum numbers for 
	the SM and the VLQs w.r.t.~the SM gauge symmetry are shown in Table~\ref{tab:quantum_numbers}.
	Note that, in the low-energy limit of the LS-SHUT, the third generation of the VLQs becomes heavy, and can therefore be integrated out below the scale of Left-Right symmetry breaking \cite{Morais:2020ypd}. %at low-scale.
	Hence, in Table~\ref{tab:quantum_numbers}, only two generations of VLQs are considered. %we only consider two generations.
	\begin{table}[H]
		\centering
		\begin{tabular}{c|c|c|c|c}
			\textbf{Field} & $\mathbf{SU(3)_\text{C}}$ & $\mathbf{SU(2)_\text{L}}$ & $\mathbf{U(1)_\text{Y}}$ & \textbf{\# of generations} \\ \hline
			$Q_\mathrm{L}$          & \textbf{3}                & \textbf{2}                & $1/3$   & 3                          \\
			$L$            & \textbf{1}                & \textbf{2}                & $-1$  & 3                          \\
			$d_\mathrm{R}$          & \textbf{3}                & \textbf{1}                & $-2/3$  & 3                          \\
			$u_\mathrm{R}$          & \textbf{3}                & \textbf{1}                & $4/3$   & 3                          \\
			$e_\mathrm{R}$          & \textbf{1}                & \textbf{1}                & $-2$              & 3                \\
			$D_\mathrm{L,R}$          & \textbf{3}                & \textbf{1}                & $-2/3$              & 2                \\
			$\phi$          & \textbf{1}                & \textbf{2}                & $1$              &  $3$
		\end{tabular}
		\caption{\label{tab:quantum_numbers} Quantum numbers of the phenomenologically relevant light SM fermion, light Higgs and VLQ sectors of the LS-SHUT model. Here, $Q_\mathrm{L}$ and $L$ are the SM quark and lepton left-handed doublets. $d_\mathrm{R}$, $u_\mathrm{R}$ and $e_\mathrm{R}$ are the right-handed down, up and lepton SM singlets, respectively. The new physics is encoded in the last two rows, with $D_{\mathrm{L,R}}$ corresponding to the VLQs and $\phi$ to the three generations of Higgs doublets containing the SM-like Higgs boson.}
	\end{table}
	
	The $\mathrm{SU(2)_L}$ doublets are defined as 
	\begin{equation}\label{eq:Doublets}
		\begin{aligned}
			Q_\mathrm{L}^i= \begin{bmatrix}
				u_\mathrm{L}\\ 
				d_\mathrm{L}
			\end{bmatrix}^i \quad
			L^i= \begin{bmatrix}
				\nu_{e_\mathrm{L}}\\ 
				e_\mathrm{L}
			\end{bmatrix}^i \quad 
			\phi^i = \begin{bmatrix}
				\phi^+ \\ \phi^0
			\end{bmatrix}^i \,,
		\end{aligned}
	\end{equation}
	%with 
	where $Q_\mathrm{L}^i$ with the flavour index $i=1,2,3$ corresponding to three generations of the left-handed SM-like quark %components 
	$\mathrm{SU(2)_L}$-doublets
	that arise from the $\bm{Q}_{\mathrm{L}}$ superfield, %$\bm{Q}_{\mathrm{L,R}}$
	whereas the SM-like lepton doublets %components, 
	$L^i$, come from the $\bm{L}$ superfield, %as shown in the
	whose interactions are determined by the superpotential \eqref{eq:Superpotential}. %We define $i=1,2,3$ as a flavour index. 
	Taking into account the representations shown in Tab.~\ref{tab:quantum_numbers}, the most general renormalisable Yukawa Lagrangian of the SM-like fermions is given by
	% as
	\begin{equation}\label{eq:Lag_quarks}
		\begin{aligned}
			\mathcal{L}_{\text{y}} = &\qty(Y^a)_{iJ}\qty(\bar{Q}_\mathrm{L})^i\qty(D_\mathrm{R})^J\phi_a + \qty(\Gamma^a)_{ij}\qty(\bar{Q}_\mathrm{L})^i\qty(d_\mathrm{R})^j\phi_a + \qty(\Delta^a)_{ij}\qty(\bar{Q}_\mathrm{L})^i\qty(u_\mathrm{R})^j\Tilde{\phi}_a + \\ &  +  \qty(\Pi^a)_{ij}\qty(\bar{L})^i\qty(e_\mathrm{R})^j\phi_a  + \text{h.c.} \,.
		\end{aligned}
	\end{equation}
	%where ``h.c.'' represents the Hermitian conjugate of the previous terms. 
	Here, we consider only two generations of the VLQs, with $J=1,2$, such that $Y$ is a $3\times 2$ Yukawa matrix, whereas 
	%the matrices 
	$\Gamma$, $\Delta$ and $\Pi$ are the Yukawa $3\times3$ textures. Note that in the main text we further consider the also valid scenario where only one generation of VLQs is light enough to be present in the low-scale particle spectrum. Due to the vector-like nature of the BSM quarks additional bilinear terms are allowed by the gauge symmetry and read
	\begin{equation}\label{eq:bilinear_mass}
		\mathcal{L}_{\text{bil}} = \qty(M_D)_{IJ}\qty(\bar{D}_\mathrm{L})^I\qty(D_\mathrm{R})^J + \qty(\Xi)_{Ij}\qty(\bar{D}_\mathrm{L})^I\qty(d_\mathrm{R})^j \,.
	\end{equation}
	Following spontaneous EW symmetry breaking, the mass Lagrangian of the quark fields takes the form
	%reads as 
	\begin{equation}\label{eq:QuarkLag_HM}
		\begin{aligned}
			&\mathcal{L}_{q,\text{SB}} =
			\frac{v_a}{\sqrt{2}}\qty(Y^a)_{iJ}\qty(\bar{d}_\mathrm{L})^i
			\qty(\bar{D}_\mathrm{R})^J +
			\frac{v_a}{\sqrt{2}}\qty(\Gamma^a)_{ij}\qty(\bar{d}_\mathrm{L})^i
			\qty(\bar{d}_\mathrm{R})^j +
			\frac{v_a}{\sqrt{2}}\qty(\Delta^a)_{ij}\qty(\bar{u}_\mathrm{L})^i
			\qty(\bar{u}_\mathrm{R})^j +\\
			&+\qty(M_D)_{IJ}\qty(\bar{D}_\mathrm{L})^I\qty(\bar{D}_\mathrm{R})^J +
			\qty(\Xi)_{Ij}\qty(\bar{D}_\mathrm{L})^I\qty(\bar{d}_\mathrm{R})^j \,,
		\end{aligned}
	\end{equation}
	with $v_a$ corresponding to the Higgs doublet VEVs. 
	%of which of Higgs doublets. 
	For the up-quark sector, the mass matrix is given as 
	\begin{equation}\label{Mass_up}
		M_u=\frac{v_a}{\sqrt{2}}
		\begin{bmatrix}
			\Delta^a_{11}&\Delta^a_{12}  &\Delta^a_{13} \\ 
			\Delta^a_{21}&\Delta^a_{22}  &\Delta^a_{23} \\ 
			\Delta^a_{31}&\Delta^a_{32}  &\Delta^a_{33}
		\end{bmatrix} \,,
	\end{equation}
	whose eigenvalues give rise to the masses of the up, charm and top quarks in the SM. On the other hand, the down-type mass matrix takes the form
	\begin{equation}\label{Mass_down}
		M_d = 
		\begin{bmatrix}
			\frac{v_a}{\sqrt{2}}\Gamma^a_{11} &\frac{v_a}{\sqrt{2}}\Gamma^a_{12} &\frac{v_a}{\sqrt{2}}\Gamma^a_{13} &\frac{v_a}{\sqrt{2}}Y^a_{11} &\frac{v_a}{\sqrt{2}}Y^a_{12} \\ 
			\frac{v_a}{\sqrt{2}}\Gamma^a_{21} &\frac{v_a}{\sqrt{2}}\Gamma^a_{22} &\frac{v_a}{\sqrt{2}}\Gamma^a_{23} &\frac{v_a}{\sqrt{2}}Y^a_{21} &\frac{v_a}{\sqrt{2}}Y^a_{22} \\ 
			\frac{v_a}{\sqrt{2}}\Gamma^a_{31} &\frac{v_a}{\sqrt{2}}\Gamma^a_{32} &\frac{v_a}{\sqrt{2}}\Gamma^a_{33} &\frac{v_a}{\sqrt{2}}Y^a_{31} &\frac{v_a}{\sqrt{2}}Y^a_{32} \\ 
			\Xi_{11} &\Xi_{21} &\Xi_{31} &\qty(M_D)_{11} &\qty(M_D)_{12} \\ 
			\Xi_{12} &\Xi_{22} &\Xi_{32} &\qty(M_D)_{21} &\qty(M_D)_{22} 
		\end{bmatrix} \,.
	\end{equation}
	In the considered model, NP effects contribute only to the down-quark sector at tree-level through a mixing with down-type VLQs. Where the lightest VLQ is denoted as $D$, whereas the heaviest VLQ is denoted as $S$.

%%%%%%%%%%%%%%%%%%%%%%%%%%%%%%%%
\section{VLQ Feynman rules}
\label{app:Feynman-Rules}
%%%%%%%%%%%%%%%%%%%%%%%%%%%%%%%%

In this appendix, we present the relevant Feynman rules, in the mass basis, that contribute to collider analysis in our work. Here, we define $g$ as the $\mathrm{SU(2)_L}$ gauge coupling, $g_s$ as the $\mathrm{SU(3)_C}$ gauge coupling, $\delta_{ab}$ as the Kronecker delta in colour space, $\delta_{ij}$ as the Kronecker delta in generation space and $\lambda$ as the well-known Gell-Mann matrices. The projector operators are defined in the standard way as $\mathrm{P_L}=(1-\gamma_5)/2$ and $\mathrm{P_R}=(1+\gamma_5)/2$. Generation indices run as $i = d, s, b, \mathrm{D},\dots$ and $j = u, c, t$, where dots denote heavier down-type VLQs ignored in the current analysis. Besides, $V_\mathrm{CKM}$ is the extended CKM matrix defined in \eqref{eq:VLQs_CKM}.
\begin{equation}\label{eq:Feynman_Rules_W_quarks}
\raisebox{-5.0em}{\includegraphics{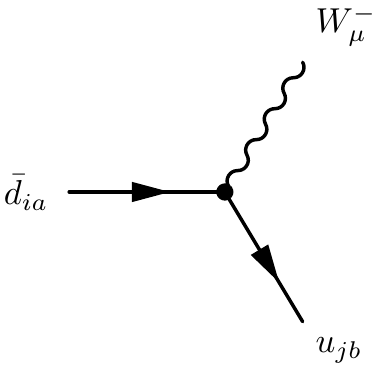}}  = \frac{ig}{\sqrt{2}} \delta_{ab} (V_{\mathrm{CKM}})_{ij} \gamma^\mu \mathrm{P_L}
\end{equation}
%%%%%%%
\begin{equation}\label{eq:Feynman_Rules_G_quarks}
\raisebox{-5.0em}{\includegraphics{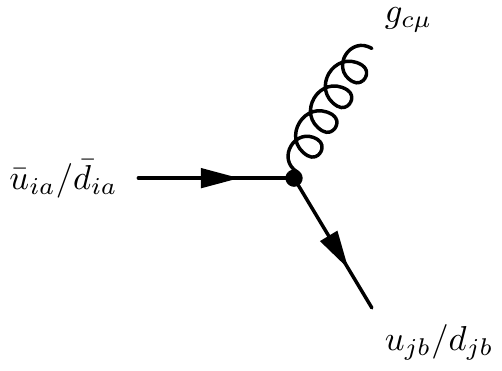}}  =-\frac{ig_s}{2} \delta_{ij}(\lambda^c)_{ab}\gamma^\mu \mathrm{P_L} -\frac{ig_s}{2} \delta_{ij}(\lambda^c)_{ab}\gamma^\mu \mathrm{P_R}
\end{equation}

\section{Angular and kinematic distributions at the ATLAS detector}
\label{app:Kinematics}
%%%%%%%%%%%%%%%%%%%%%%%%%%%%%%%%%%%%%%%%%%%%%%%%%%%%%%%%%%%%%%%%%%%%%%%

%%%%%%%%%%%%%%%%%%%%%%%%%%%%%%%%%
\begin{figure}[htb!]
	\centering
	\includegraphics[width=\textwidth]{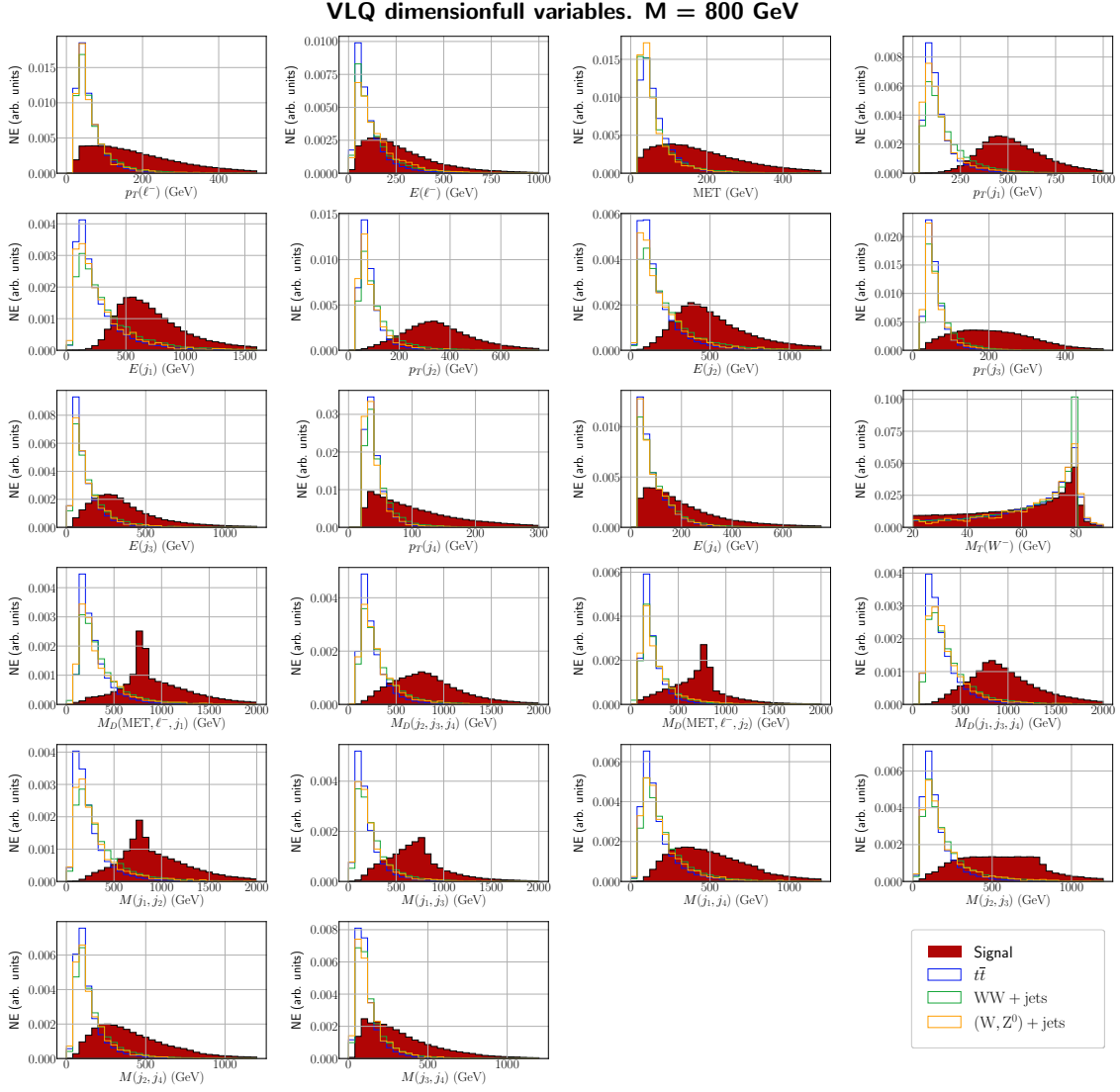}
	\caption{The dimensionfull kinematic distributions (in units of GeV) for pair-production of 800 GeV VLQs. From top to bottom and left to right, we have $p_T(\ell^-)$, $E(\ell^-)$, MET, $p_T(j_1)$, $E(j_1)$, $p_T(j_2)$, $E(j_2)$, $p_T(j_3)$, $E(j_3)$, $p_T(j_4)$, $E(j_4)$, $M_T(W^-)$, $M(W^+_{jj})$, $M_D(\mathrm{MET},\ell^-,j_1)$, $M_D (j_2,j_3,j_4)$, $M_D(\mathrm{MET},\ell^-,j_2)$, $M_D (j_1, j_3, j_4)$, $M(j_1,j_2)$, $M(j_1,j_3)$, $M(j_1,j_4)$, $M(j_2,j_3)$, $M(j_2,j_4)$ and $M(j_3,j_4)$. The $y$-axis represents normalised events, and we select 30 bins for both the signal and the background.}
	\label{fig:Dimensionfull-vars}
\end{figure}
%%%%%%%%%%%%%%%%%%%%%%%%%%%%%%%%%
%%%%%%%%%%%%%%%%%%%%%%%%%%%%%%%%%
\begin{figure}[htb!]
	\centering
	\includegraphics[width=\textwidth]{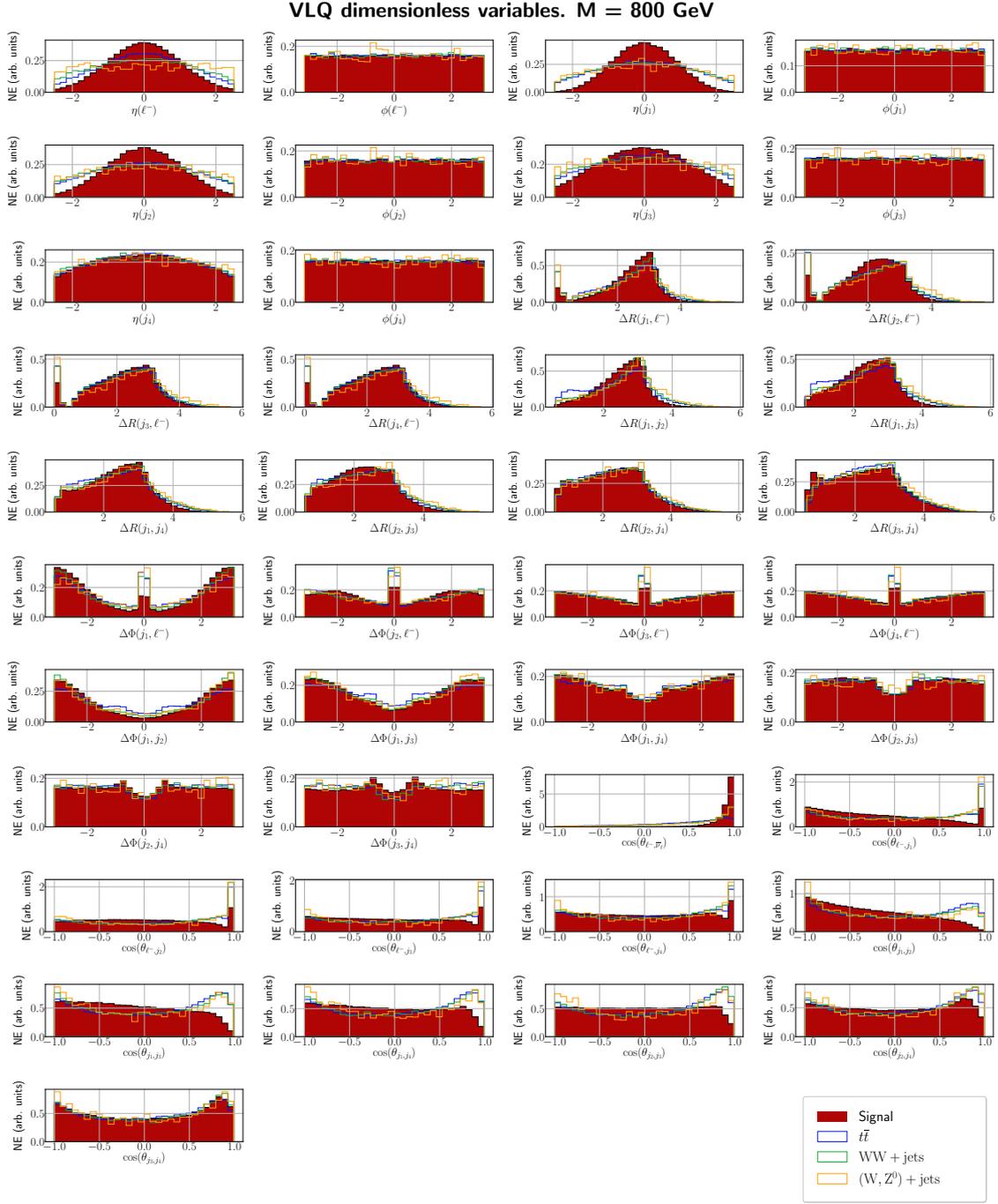}
	\caption{The dimensionless angular distributions for pair-production of 800 GeV VLQs. From top to bottom and left to right, we have  $\eta(\ell^-)$, $\phi(\ell^-)$, $\eta(j_1)$, $\phi(j_1)$, $\eta(j_2)$, $\phi(j_2)$, $\eta(j_3)$, $\phi(j_3)$, $\eta(j_4)$, $\phi(j_4)$, $\Delta R (j_1,\ell^-)$, $\Delta R (j_2,\ell^-)$, $\Delta R (j_3,\ell^-)$, $\Delta R (j_4,\ell^-)$, $\Delta R (j_1,j_2)$, $\Delta R (j_1,j_3)$, $\Delta R (j_1,j_4)$, $\Delta R (j_2,j_3)$, $\Delta R (j_2,j_4)$, $\Delta R (j_3,j_4)$, $\Delta \Phi (j_1,\ell^-)$, $\Delta \Phi (j_2,\ell^-)$, $\Delta \Phi (j_3,\ell^-)$, $\Delta \Phi (j_4,\ell^-)$, $\Delta \Phi (j_1,j_2)$, $\Delta \Phi (j_1,j_3)$, $\Delta \Phi (j_1,j_4)$, $\Delta \Phi (j_2,j_3)$, $\Delta \Phi (j_2,j_4)$, $\Delta \Phi (j_3,j_4)$, $\cos(\theta_{\ell^-,\bar{\nu}_\ell})$, $\cos(\theta_{\ell^-,j_1})$, $\cos(\theta_{\ell^-,j_2})$, $\cos(\theta_{\ell^-,j_3})$, $\cos(\theta_{\ell^-,j_4})$, $\cos(\theta_{j_1,j_2})$, $\cos(\theta_{j_1,j_3})$, $\cos(\theta_{j_1,j_4})$, $\cos(\theta_{j_2,j_3})$, $\cos(\theta_{j_2,j_4})$ and $\cos(\theta_{j_3,j_4})$. The $y$-axis represents normalised events and we select 30 bins for both the signal and the background.}
	\label{fig:Dimensionless-vars}
\end{figure}
%%%%%%%%%%%%%%%%%%%%%%%%%%%%%%%%%

\cleardoublepage
\bibliographystyle{JHEP}
\bibliography{bib}

\end{document}